# Rethinking Resource Allocation in Science

Johan Bollen* [1,4], Stephen Carpenter [2], Jane Lubchenco [3], and Marten Scheffer* [4]

1: School of Informatics,  Computing, and Engineering, Indiana University, Bloomington IN
2: Center for Limnology, University of Wisconsin-Madison
3: Department of Integrative Biology, Oregon State University
4: WU Environmental Sciences, Wageningen University, Wageningen, The Netherlands

*: To whom correspondence should be addressed; E-mail:  jbollen@indiana.edu, marten.scheffer@wur.nl

US funding agencies alone distribute a yearly total of roughly $65B dollars largely through the process of proposal peer review: scientists compete for project funding by submitting grant proposals which are evaluated by selected panels of peer reviewers. Similar funding systems are in place in most advanced democracies. However, in spite of its venerable history, proposal peer review is increasingly struggling to deal with the increasing mismatch between demand and supply of research funding.

**A costly system**. The most conspicuous problem with the current system is the cost associated with the time spent on writing, processing, and reviewing project proposals. For instance, it is estimated that European researchers collectively spent about €1.4 billion worth of time to submit unsuccessful applications to the Horizon 2020 program, a sizable proportion of the €5.5 billion distributed. Meanwhile, Australian researchers are estimated to collectively spend three centuries a year writing, submitting, and reviewing project proposals (*1*). Of course, this time is not entirely lost. Writing and reviewing proposals helps to articulate one's vision, but is that worth the extraordinary amount of time spent? Does the system allocate science funding in the most effective manner so that society receives the greatest possible return on investment? Intuitively, one would think so, but the capacity of peer review to sort out the most productive proposal is in fact surprisingly low. For instance, analysis of 102,740 funded NIH grants found almost no relationship between review scores and the resulting scientific output (*2*).

So, should we simply skip the proposal submission and review machinery? We could, for example, give all tenured researchers an equal share of available funding. An analysis of the Natural Science and Engineering Research Council Canada (NSERC) statistics shows that preparing a grant application costs approximately $40,000 (Canadian). This is more expensive than simply giving every qualified investigator a direct baseline discovery grant of $30,000 (*3*). On the other hand, not all scientific work is equal. Some scientists do conduct research that is more promising and some efforts inherently require greater resources. Awarding the same amount of baseline funding to every researcher is therefore not an optimal strategy. Furthermore, an equal distribution may not meet societal or programmatic needs.

Could we redesign our funding system in a way that reduces both excessive costs and low accuracy, while ensuring the fundamental needs of society are being met? Here, we suggest a redesign based on two simple starting points:

**A) Fund people instead of projects.** This principle obviates the need for project proposal writing, reviewing, and management. At the same time, it likely improves reliability because it is based on the evaluation of the comprehensive merits of individual scientists rather than a single project proposal. Indeed, a comparative study suggests that a person-based funding system results in more high-quality scientific output (*4*).

**B) Leverage the wisdom of the crowd.** There is strong evidence that large groups can collectively make better decisions than small teams of specialists as long as decisions are made independently and the groups are sufficiently diverse (*5*).

To see how a system based on these two principles may work for fund allocation, consider the following two-step procedure:

1. Every participating scientist receives an equal portion of all available funding as their base starting budget.
2. Each participant anonymously donates a fixed percentage (say 50%) of their funding to other, non-affiliated scientists.

This is repeated each funding round.

It is important to note that each scientist must distribute a percentage of *everything* they received in the previous round, i.e. the base funding plus what they previously received from other scientists. For example, suppose that a scientist receives the base amount of $50,000 and received $150,000 from other researchers. The total received is $200,000 of which 50%, i.e $100,000, needs to be donated to other scientists. The scientist retains a total of $100,000. Since every scientist participates, funding circulates through the community converging over time to funding levels that all scientists have *collectively, yet independently* determined. Importantly, scientists that receive more funding than others also become the more significant funders in the system. This self-organized weighting resembles the mathematical technique of power iteration used to converge on stationary probability distributions of web page relevancy (*6*).

To speed up the process, the initial manual funding selections can be algorithmically carried forward until a convergence criterion is reached. Another option is to have a two-phase donation process. A first round is followed by the publication of funding numbers and subsequently a second donation round. Regardless of the specific implementation details, the system converges on a distribution of funding that reflects all information in the scientific community with a minimum investment of time and effort.

**Challenges**
While the basic principle is simple and transparent, its practical implementation requires some additional considerations. First of all, we have to decide who can participate in this system. For example, the system could involve everyone with an academic appointment at an accredited institution. Second, like the current proposal peer review system, conflicts of interest must be vigorously prevented. A well-designed automated approach may effectively eliminate most problems.  For instance, co-authorship and shared affiliations can be automatically detected from scientific information databases. Also, algorithms may efficiently detect fraudulent reciprocal donation loops or cartels which should be forbidden and penalized.

Funding agencies will naturally play a central role in the development,  application, and refinement of the proposed system.  For instance, SOFA could be set-up to run within specific domains, subdomains, or even smaller topic areas (e.g. Chemistry, environmental chemistry, or marine chemistry). This allows funding agencies and policy makers to set budgets according to programmatic priorities. Stable funding for expensive infrastructure and long-term contracts could continue to be allocated by the existing funding system.  However, staying closer to the new approach, researchers could also be allowed to put up large common projects or infrastructures as "super-nodes" for funding in SOFA. It may also be convenient to provide some generic options such as "redistribute my funding equally to all female scientists" or "scientists younger than 30 years old". These and other elaborations may further ensure a reliable and balanced system. Clearly, a cautious approach is needed, requiring a transparent multi-disciplinary team effort that involves the funding agencies for designing, monitoring, and evaluating pilot projects that pave the way for a larger scale implementation.

**Opportunities** While there are obvious challenges and uncertainties in implementing such a novel approach, there are also opportunities that go beyond solving the excessive overhead and unreliability of the current system. There are at least four commonly recognized issues that can be addressed in one stroke:
1) *Systematic biases* with regard to ethnicity or gender can be objectively measured and mitigated. For instance, a bias towards funding women may be corrected by raising the funding to each female scientist by a fixed percentage.
2) *Excessive inequality* in funding can be controlled by tuning the mandatory donation fraction. Simulations suggest that a 50% donation fraction results in funding inequality that approximates that of the current system (*6*). By contrast a very small donation fraction will result in a highly egalitarian distribution since scientists simply retain most of their base funding (*6*).
3) *Newcomers* always receive the guaranteed base fund with no obligations to spend excessive time in applying for project funding. A reduced mandatory donation fraction for early career scientist could strengthen their position even further.

*The ivory tower* effect could be reduced by letting a percentage (say 10%) of the funds be distributed by the public allowing for transparent input with respect to societally desirable research directions. This would in addition stimulate researchers to communicate their ideas to the public.

**Risks, barriers, and bridges**
The proposed system would immediately save billions of dollars that are now spent on proposal submission and reviewing. While it has the potential to solve a range of broadly felt issues with our present system of science funding, it remains impossible to foresee all consequences. A donation system may lead to higher well-being among researchers (*7*) than the present competition-oriented model, but at the same time the crowd-based aspect will reward those who most openly communicate their work and plans, encouraging "salesmanship" at the cost of thoughtfulness.

A central challenge will be to ensure that the system remains responsive to societal needs. Will the wisdom of the crowd converge to priorities and objectives that meet societal needs? Our proposal includes the ability of policy-makers to direct funding to particular domains and constituencies. Clearly, funding agencies will remain uniquely positioned to provide guidance and know-how for bridging societal and scientific objectives. Government program managers would continue to be highly engaged in the process, but their role would shift toward designing useful classification structures, for instance defining sub-domains, and managing crowd-based assessments within those domain (rather than the laborious task of evaluating scientific excellence). Instead of directing funds to scientists, the agencies would work collaboratively with scientists and decision-makers, to leverage shared resources to support both scientific excellence and programmatic obligations.

Fortunately, implementation is not an "all or nothing" matter. One could run small-scale trials with fractions of the national research budget alongside the existing system. This might in fact soon be realized in The Netherlands where the Dutch parliament approved a motion directing the national science funding agencies to experiment with new models of funding allocation. Such tests afford the opportunity to conduct repeated cycles of evaluation that can inform gradual improvement to the system as it is being scaled up.

The funding model we propose may seem a potentially disruptive innovation. However, society can no longer afford to lose billions in a complex and costly machinery with unclear performance. The present system has served us well for over half a century. It may now be perceived as 'tested-and-proven', but we have come to a point that incremental adjustments seem unlikely to repair its broadly recognized shortcomings. The situation we face may be an example of how scaling-up can sometimes lead to fundamentally unsustainable overhead as observed in systems ranging from businesses (*8*) to societies (*9*). A carefully planned experiment with a Self Organized Fund Allocation system may provide a bridge to more efficient and reliable alternatives.

**Acknowledgements:**
This manuscript formed the basis of a 2018 Nature Worldview (10) editorial. The authors express their gratitude to Kate Coronges and Alessandro Vespignani of the Network Science Institute, Northeastern University, Boston MA for their tremendously constructive feedback and comments that helped us to significantly improve this manuscript. We also want thank the organizers and participants of the national workshop on increasing grant submission pressure ("Aanvraagdruk") and improving NOW grant request procedures that was organized by Netherlands Organization for Scientific Research (NWO) on April 2017 (https://www.nwo.nl/beleid/nwo+werkconferenties+2017/nationale+werkconferentie)

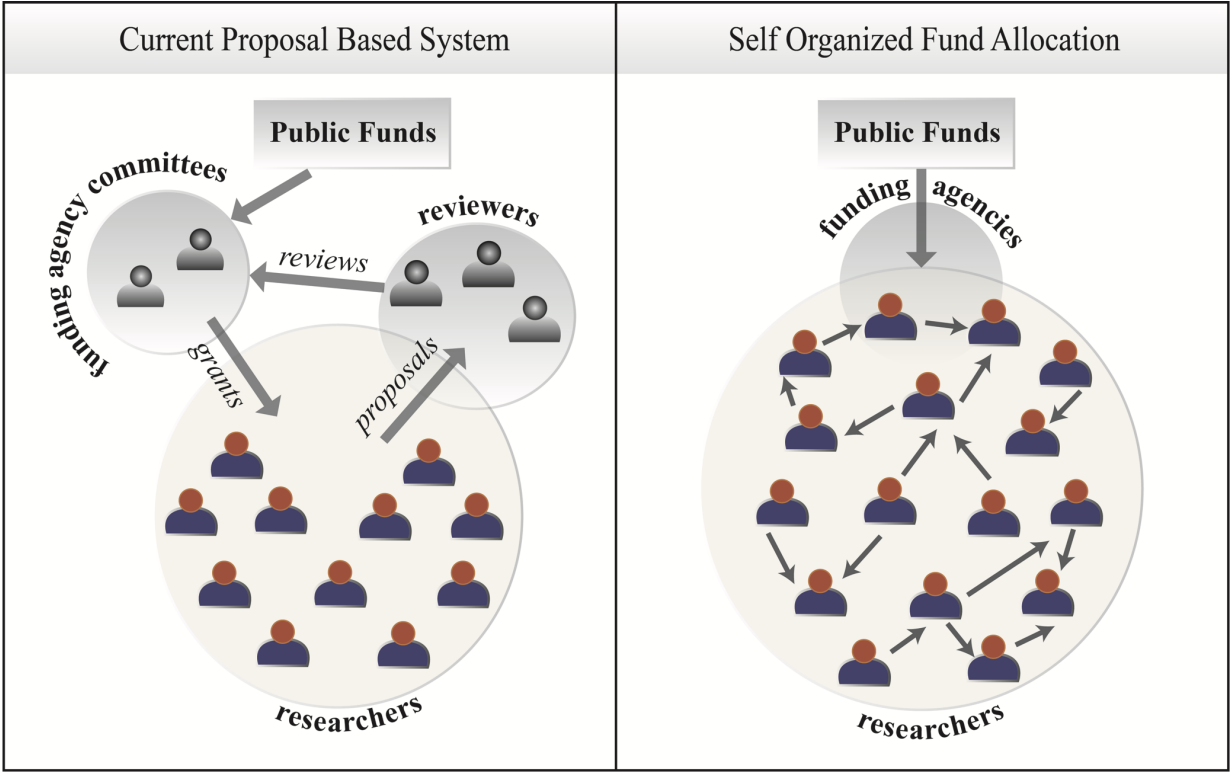

*Figure 1* Schematic comparison of the current fund allocation model compared to Self-Organized Fund Allocation.